\documentclass{ws-procs975x65}

\usepackage{latexsym}
\usepackage{amssymb}
\usepackage{amsmath}
\usepackage{float}
\usepackage{indentfirst}
\usepackage{array}
\usepackage{xspace}

\usepackage[unicode=true]{hyperref}
\usepackage{amsfonts}
\usepackage{graphicx}
\usepackage{epic}
\usepackage{eepic}
\usepackage{psfrag}
\usepackage[mathcal]{euscript}

\def\be{\begin{equation}}
\def\ee{\end{equation}}
\def\ba{\begin{eqnarray}}
\def\ea{\end{eqnarray}}

\def\lb{\label}
\def\H{{\cal H}}
\def\e{{\rm e}}

\begin{document}
\title{SYNCHROTRON RADIATION AND PAIR CREATION \\OF MASSLESS CHARGES IN MAGNETIC FIELD}
\author{D.\,V.\,GAL'TSOV }
\address{\parbox{10cm}{\noindent\rule{0cm}{0.4cm}{} Faculty
of Physics, Moscow State University, 119899, Moscow, Russia.}  \\
\email{galtsov@phys.msu.ru}}
\begin{abstract}
We study the massless limit in  synchrotron radiation and one-photon
pair creation in magnetic field. In this limit Schwinger critical
field  $H_0=m^2c^3/(e\hbar)$  tends to zero, so two characteristic
quantum parameters $\eta=H/H_0,\;\chi=\eta E/mc^2$ are infinite, and
the standard approximations used in analytical calculations fail.
Applying Schwinger's proper time methods we derive simple
expressions for synchrotron radiation spectra emitted by massless
charges of spins $s=0,\,1/2$ and the pair creation probability
distribution in the quasiclassical (high Landau levels) regime
exhibiting simple scaling properties and possessing universal
spectral shapes.
\end{abstract}
\bodymatter
\section{Introduction}\label{intro}
It could be naively thought that the ultrarelativistic limit in the
radiation problem is identical with the limit of zero mass of the
radiating charge. That this is not so simple is clear already
looking at the classical Lienard formula \cite{LL} for radiation
power of the relativistic massive charge moving with the transverse
proper acceleration $a$ and the velocity $v$ (in the units $c=1$):
$P =2/3\, e^2 a^2 (1-v^2)^{-2} $, which diverges in the massless
limit $v \to 1$. At the same time, it was argued that massless
charges would not  radiate at all
\cite{Lechner:2014kua,Kosyakov:2007pm}, an assertion, which could
partly explain their experimental non-observation. But an infinitely
growing radiation power is unacceptable since  the particle can not
emit the energy greater than its own energy, so the quantum nature
of radiation must be invoked. On the other hand, the absence of
radiation contradicts to Bohr quantum principles, if the charge
occupies the excited energy states. So, in some sense, taking the
ultrarelativistic limit does not commute with quantization.

Passing  to  the most interesting   case of radiation in magnetic
field, one finds that  the ultrarelativistic limit $E\gg m$ may be
essentially different depending on whether the dimensionless quantum
recoil parameter $\chi=eHE/m^3$ is small or large (we will use units
$\hbar=c=1$ unless stated explicitly). In the first case the second
dimensionless parameter $\eta=eH/m^2$, defining the ratio of the
Landau spacing  to the mass in the energy spectrum
$E=\sqrt{m^2+eH(2n+1)}$ (for zero spin) is also small, and the
ultrarelativistic limit is mostly classical. This is typical for the
laboratory situation. The ultra-quantum case $\chi\gg 1$ and
$\eta\ll 1$ is also well-known: the corresponding conditions are
commonly realized in pulsars with strong magnetic fields. But in the
strictly massless case {\em both} parameters are infinite, so
special consideration is required. In the existing literature one
can find the double Taylor expansions of the radiation power in
terms of $\chi,\; \eta$, and some complicated formulas for arbitrary
$\eta$. Here we  derive quite simple expressions relevant to the
case of both parameters infinite which have simple scaling features.
In this case most of radiation from the highly excited Landau levels
$1\ll n\leq 10^7$ (the upper restriction ensuring the charge energy
spectrum to be quasi-discrete) is emitted in the transitions $n\to
n'$ with large $n'$. Since the standard approximations used in the
massive case fail, one needs to redo the derivation. But after all,
it turns out that that the main contribution can still be extracted
from the massive ultra-quantum limit $\chi\gg 1$ and $\eta\ll 1$,
provided the above restrictions on $n$ holds.

Although our primary goal is to clarify the double limit
$\chi\to\infty,\; \eta\to \infty$ of the quantum synchrotron
radiation spectrum  of the {\em massive} charges, it is worth saying
a few words about validity of QED for truly {\em massless} ones.
Attempts to find theoretical arguments explaining non-existence of
such charges have long history \cite{Vaks:1961,Case:1962zz}. In
perturbation theory they cause  collinear, or massless,
singularities \cite{Kinoshita:1962ur,Lee:1964is} of the Feynmann
diagrams occurring when the photons  are emitted from the massless
external legs with the momentum parallel to the charge momentum
\cite{Weinberg:1965nx}. This problem, however, can be cured just
ramifying the calculations \cite{Kinoshita:1962ur,Lee:1964is}. Also,
the arguments were presented that free massless charges would be
screened by vacuum
polarization~\cite{Fomin76,Gribov:1981jw,Morchio:1985re}, the
phenomenon which was interpreted as charge confinement in the
massless QED. The corresponding length, however, is quite large, so
the smaller scale processes, like pair creation, are not forbidden.
Anyway, independently of the consistency problem of perturbative
QED, the situation changes  once classical magnetic field is added
non-perturbatively. Such {\em magnetized} massless QED looks to be a
consistent theory leading to interesting predictions is the
low-energy region \cite{Ferrer:2012pb}. In \cite{Gal'tsov:2015cla}
we have considered synchrotron radiation from massless charges in
magnetized scalar QED showing that radiation do exist and has the
universal spectral shape depending on neither the magnetic field nor
the energy. Here we briefly review this calculation and give similar
formulas for the spin 1/2 charge and for the one-photon massless
pair creation.
\section{Universality of radiation spectra from massless charges }
 Quantum theory of synchrotron radiation has different
formulations. Historically the first one  was based on  exact
solutions of the Klein-Gordon and Dirac equations in magnetic field
\cite{Sokolov1} (for more recent review see \cite{Bor}). Later on,
Schwinger and Tsai applied the ``proper time'' method to calculate
the one-loop mass operator of  massive charges with $s=0,\,1/2$ in
the constant magnetic field $H$, its imaginary part gives the total
rate of synchrotron radiation \cite{Schwinger:1973,Tsai:1974cc}.  We
have repeated similar calculations for  charged particles of
strictly zero mass. For $s=0$ the main steps go as
follows~\cite{Gal'tsov:2015cla}.

The mass operator for the complex scalar field  in Schwinger's
operator notation reads:
 \be
 M=ie^2\int\left[(\Pi-k)^\mu\frac1{k^2}\frac1{(\Pi-k)^2}(\Pi-k)_\mu\right]\;
 \frac{d k}{(2\pi)^4}\;
 \;-M_0\,,
 \ee
where $
 \Pi_\mu=-i\partial_\mu-e A_\mu,\;\,A_\mu
$ stands for the constant magnetic field, and $M_0$ is the
subtraction term needed to ensure vanishing of $M$ and its first
derivative with respect to $ \Pi^2$ at $\Pi^2=0$. Exponentiating the
propagators,
 \be
\frac1{k^2}\frac1{(\Pi-k)^2 }=-\int_0^\infty
sds\int_0^1\e^{-is\H}\,,\quad \H=(k-u\Pi)^2-u(1-u) \Pi^2\,,
 \ee
and  replacing the $k$-integration by averaging over the eigenstates
of the coordinate $\xi^\mu$ of the fictitious particle, canonically
conjugate to $k_\mu$,  one obtains:
 \be
M=ie^2\int_0^\infty sds\int_0^1 du   \langle
\xi\big|(\Pi-k)^\mu\e^{-is\H} (\Pi-k)_\mu\big|\xi\rangle \;\; -M_0
\,.
 \ee
The quantity $\H$ is then treated as the Hamiltonian,  and the
averaging is performed in the Heisenberg picture passing to
$s$-dependent operators $k(s), \xi(s), \Pi(s)$, which can be found
exactly in terms of $H$. Performing  calculations in the massless
case one arrives at
 \be\lb{Mint}
M=\frac{e^2}{4\pi}\int_0^1 du\int_0^\infty
\frac{ds}{s}\left[\e^{-i\psi}\Delta^{-1/2}\left(E^2\Phi_1+4ieH\Phi_2+
 i \Phi_3/s\right)-2i/s\right]\,,
  \ee where $\Phi_{1,2,3}$ and $\Delta$ are some functions of $u$ and $s$
  \cite{Gal'tsov:2015cla},
  the phase is $\psi=(2n+1)[\beta-(1-u)x],\; x=eHsu$ and
 $
  \beta= \arctan\left(\cot
x+  u /  [x(1-u)] \right)^{-1}\,. $ This expression is valid for all
Landau levels $n$. In the case $n\gg 1$ the integrals can be
evaluated expanding all $x$-dependent quantities in power series.
Indeed, the main contribution to the integral over $x$ comes from
the region where the  phase $\psi(x,u)\lesssim 1$, in which $\beta$
for $x\ll 1$ can be approximated as $ \beta \approx
(1-u)x+u(1-u)^2x^3/3\,,
 $
so that
 \be \lb{psix}
\psi \approx  3 s^3  (eHE)^2 u^4(1-u)^2/3\,. \ee For large $n$,
apart form the narrow regions around the limiting points of $u$, $
u>n^{-1}\,,\quad 1-u>n^{-1/2},\, $ which give negligible
contributions under conditions specified later on, the essential
 domain of $x$ is
$
  x\lesssim  n^{-1/3}\,.
$ Therefore we use (\ref{psix}) in the exponent, expanding the rest
of the integrand in powers of $x$:
 \begin{align}
&\Delta^{-1}\approx 1+u(4-3x) x^2/3\,, \quad
 \Phi_1\approx \left(8- {32u}/{3} +13 u^2/3-  u^3 \right)(1-u)x^2\,,\\
&\Phi_2\approx 2(1-u+u^2)x\,,\quad \Phi_3\approx 2-(4- {10
u}/{3}+u^2)x^2\,.
  \end{align}
The leading contribution comes from $\Phi_1$, while in $\Phi_3$ one
has to keep only the zero order term. Introducing the decay rate  $
\Gamma=-\frac1{E}\;{\rm{Im}}\, M\,,
$
 we obtain
 \be
 \Gamma=\frac{e^2}{4\pi E}\int_0^1 du \int_0^\infty
 \frac{dx}{x}\left(E^2\Phi_1\sin\psi\;
 +2\frac{eHu}{x}(1-\cos\psi)\right)\,,
\ee where one has to use (\ref{psix}) for $\psi$. Integrating over
$x$ and $u$, we obtain  the total decay rate exhibiting the
following scaling law:
 \be\lb{Ga}
 \Gamma=\frac{4e^2}{9E}\Gamma\left( 2/3\right)\left(3eHE\right)^{2/3}\,.
 \ee

With account for radiative decay, the energy levels  are no longer
stationary, but quasi-stationary, provided the level spacing $\Delta
E=E_n-E_{n-1}\approx (eH/2n)^{1/2}$ is  larger than $\Gamma/2$. This
leads to an upper bound for the Landau level $n$ : $
 2n<  {81}/ [{2\left[\Gamma\left( 2/3\right)\right] } {\alpha }]^3\approx
   10^7
  \,,
$ where $\alpha=1/137$.
 Thus our calculation is consistent under the following conditions
 for the energy:
$
 eH\ll E^2< 10^7 eH
  \,.
$

The real part of the mass operator gives rise to magnetically
induced mass square $\delta m^2$. This quantity is finite, while the
corresponding linear correction $\delta m$ diverges for $m=0$ in
view of the variation formula $\delta m=\delta  m^2/(2m)$. Keeping
the leading terms in the real part of (\ref{Mint}), we find:
 \be
\delta m^2= \frac{e^2}{4\pi }\int_0^1 du \int_0^\infty
 \frac{dx}{x}\left(\cos\psi\;
 \Phi_1+2 \sin\psi\frac{eHu}{x}\right) =
 \frac{4e^2\;\Gamma\left(2/3\right)}{9\sqrt{3}}\;\left(3e
HE\right)^{2/3} \,.
 \ee
This expression is essentially non-perturbative in $\alpha$.

\begin{figure}[tb]
\begin{center}
\begin{minipage}[t]{0.48\linewidth}
\hbox to\linewidth{\hss%
\psfrag{1}{{$\mu t$}} \psfrag{2}{\Large{$\frac{\psi}{\mu}$}}
\psfrag{3}{{$\psi\propto e^{-\mu t/4}$}}
  \includegraphics[width=0.95\linewidth,height=0.7\linewidth]{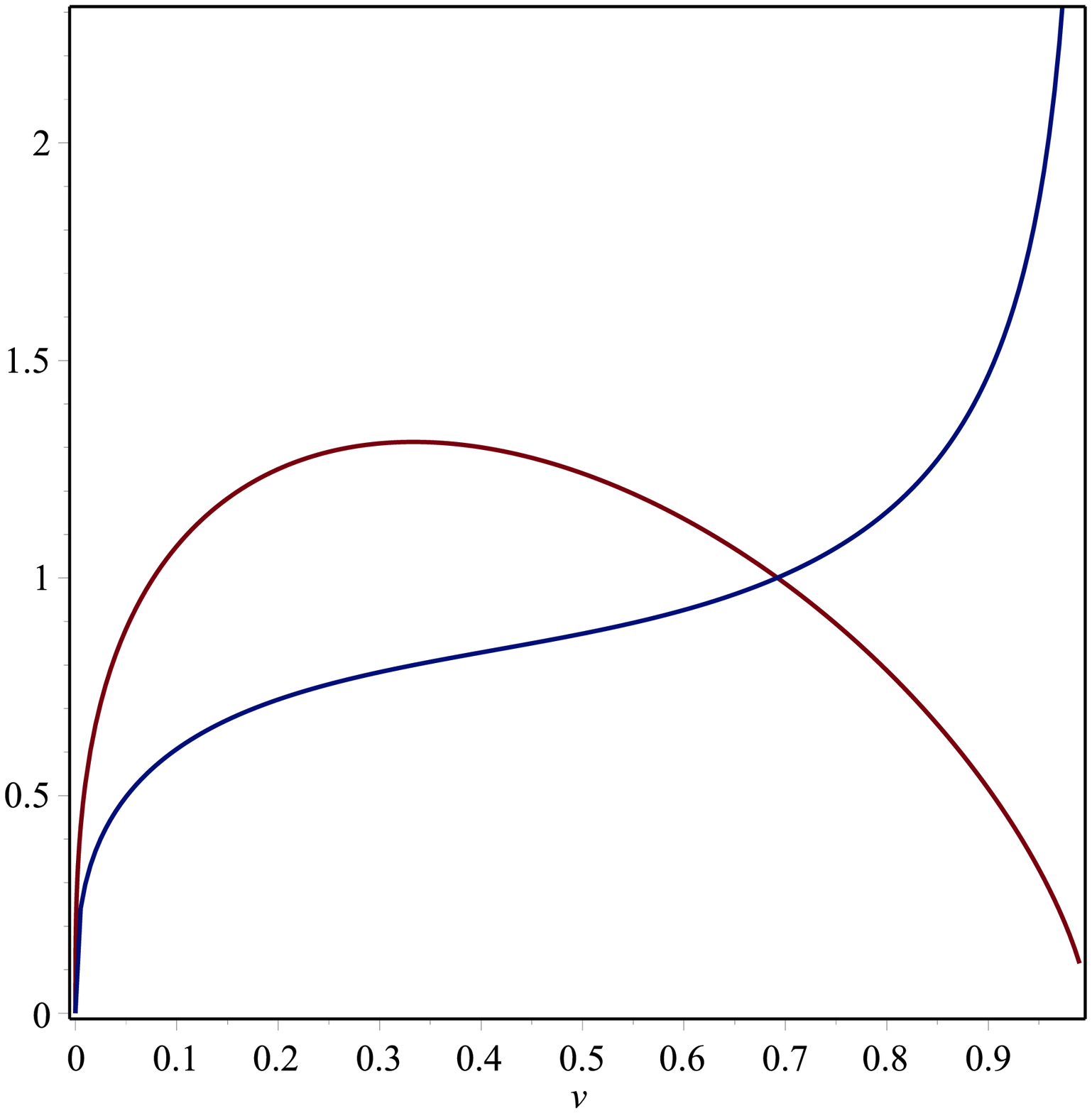}
\hss} \caption{\small   Universal synchrotron radiation spectra
${\cal P}_s(v),\; v=\hbar\omega/E,$ for massless charges of spin
$s=0$ (regular curve) and $s=1/2$ (divergent as $\to 1$).}
\label{F1}
\end{minipage}
\hfill
\begin{minipage}[t]{0.48\linewidth}
\hbox to\linewidth{\hss%
\psfrag{1}{{$\mu t$}} \psfrag{2}{\Large{$\frac{H}{\mu}$}}
  \includegraphics[width=0.95\linewidth,height=0.7\linewidth]{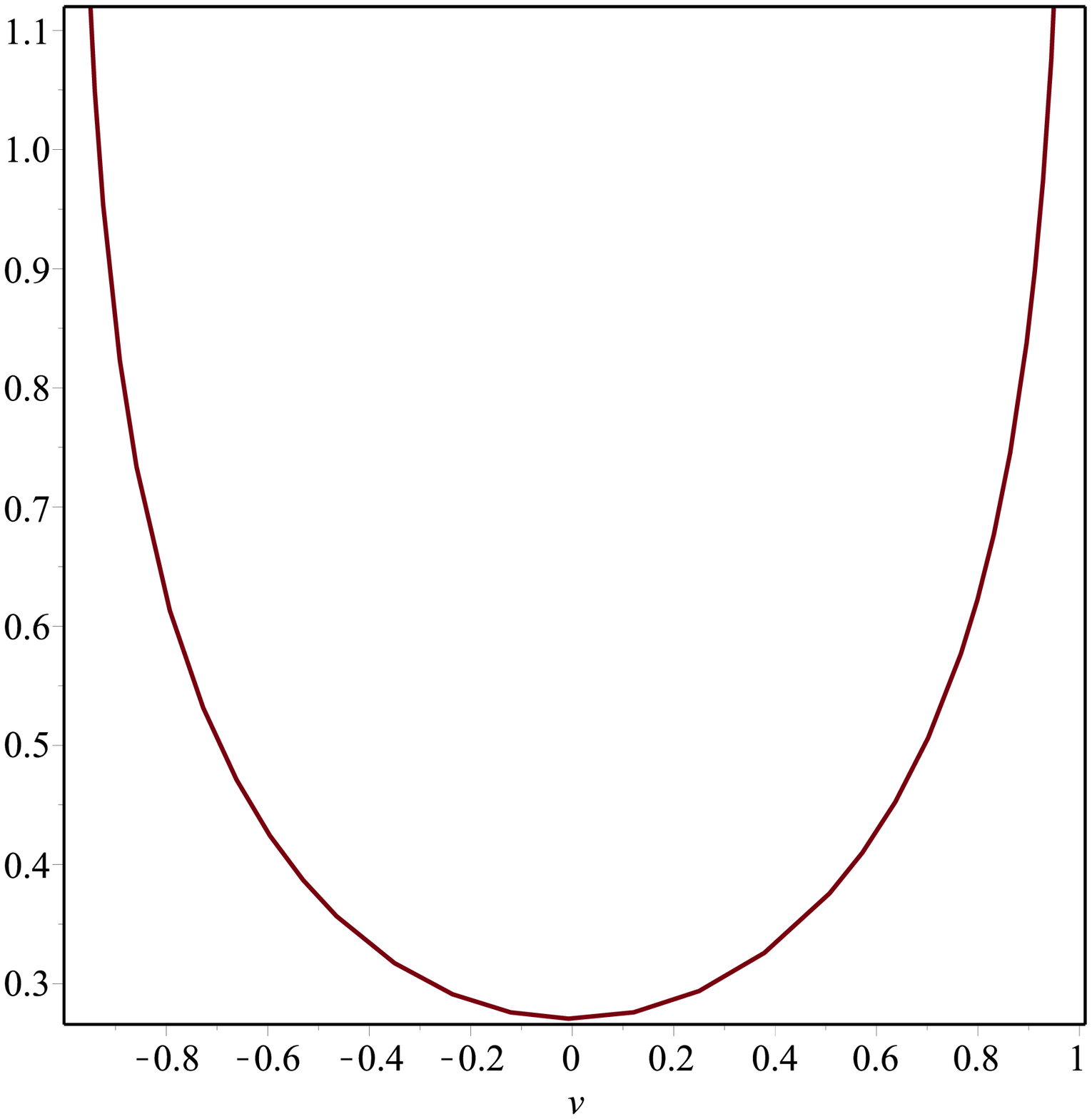}
\hss} \caption{\small   Differential probability of the one-photon
 massless $s=1/2$ pair creation as function of
 $v=(E_--E_+)/\hbar \omega$.}
\label{F2}
\end{minipage}
\end{center}
\end{figure}

To get the spectral power of radiation $P(\omega)$ one has to
perform spectral decomposition in the mass operator. Denoting
$v=\omega/E$, one obtains in the same approximation:
\cite{Gal'tsov:2015cla}
 \be
P(\omega)= \frac{e^2v}{4\pi E}\int_0^\infty
  \left(E^2(8-v^2)(1-v)^2x\sin\psi\;
 +\frac{eHv}{x^2}(1-\cos\psi)\right)dx\,,
 \ee
Evaluating the integral over $x$ and restoring $\hbar$, we get
 \be\lb{speless}
P(\omega)=\frac{2e^2\;\Gamma\left(2/3\right)}{27\hbar
E}\,\left(3e\hbar HE\right)^{2/3}{\cal P}_0\left(
{\hbar\omega}/{E}\right)\,,
 \ee
 where  the normalized spectral function is introduced
  \be \lb{spenorm} {\cal
P}_0\left(v\right)=\frac{27}{2\pi\sqrt{3}}\;v^{1/3}(1-v)^{2/3}\,,\qquad
\int_0^1{\cal P}_0\left(v\right)dv=1\,,
 \ee
which does not depend on any parameter. This spectrum, shown in
Fig.~\ref{F1}, has maximum  at $ \hbar \omega_{\rm max}= E/3,$ the
average photon energy being
 $
\langle\hbar\omega\rangle= 4 E/9 \,.
 $
Thus, radiation of the massless charge is  essentially quantum,
whose power diverges in the formal limit $\hbar\to 0$. Universality
of the spectrum means that  even in a weak magnetic field the
massless charge converts its energy into quanta of the same order of
energy.

 In the case of spin 1/2 calculations are essentially similar and
lead to the following expression for the spectral power: \be
 P(\omega)= \frac{e^2 E v}{ \pi }\int_0^\infty(v^2-2v+2)(1-v)
 x\sin\psi dx= \frac{e^2
 \Gamma(2/3)2^5\left(3eHE\right)^{2/3}}{3^5}{\cal P}_{1/2}\,,
  \ee
  where the normalized universal spectral function reads
\be \lb{spefnorm} {\cal
P}_{1/2}\left(v\right)=\frac{81\sqrt{3}}{64\pi}\;
v^{1/3}(1-v)^{-1/3}(v^2-2v+2) \,.
 \ee
At the upper limit $v \to 1$ the spin 1/2 spectral power has an
integrable divergence.  Both $s=0,\, 1/2$ spectra are plotted in
Fig.1. The low-frequency limits are identical for both spins and
coincide with the classical spectrum ($\hbar$ disappears):
  \be \lb{specla}
   P_{\rm cl}(\omega)=e^2\frac{ 3^{1/6}}{ \pi}\Gamma(2/3)\left(\frac{\omega}{\omega_H}
   \right)^{1/3}\omega_H\,,\qquad \omega_H=\frac{eH}{E}\,.
 \ee
This power-low dependence exhibits ultraviolet catastrophe (no
high-frequency cutoff), which is cured in quantum theory.

One can also investigate the case of vector massless particles,
$s=1$, but then the result is infinite: magnetized vector QED fails
to describe radiation from massless vector charges. This could be
expected in view of the results of Case and Gasiorovich
\cite{Case:1962zz}, who gave  the arguments that electromagnetic
interaction of massless charged particles with spin  one and higher
is controversial. 

\section{One-photon pair creation}
Consider now the one-photon pair creation $\gamma(k) \to
e_-(p)+e_+(p')  $ where the states of the massless pair in magnetic
field are specified by two quantum numbers $p_z, \, n$. This process
is possible due to non-conservation of the transverse momentum ---
only the energy and the longitudinal momentum are conserved in
magnetic field:
 \be
 \omega=E_-(p_z,  n)+E_+(p'_z,  n')\,,\qquad p_z+p'_z=k_z.\,
   \ee
Ignoring refraction due to vacuum polarization, we consider  the
on-shell $k^2=0$ photon propagating orthogonally to the magnetic
field, $k_z=0$. The probability of the pair creation was calculated
using the solutions of the Dirac equation in magnetic field by
Klepikov \cite{Klepikov} and as the imaginary part of the photon
polarization operator  in \cite{TE}. Using the last method in the
massless case, we obtain for the differential probability a simple
formula, shown in Fig. 2,
 \be
\frac{dW}{dv}=\frac{e^2
2^{2/3}\Gamma(2/3)(3eHE)^{2/3}}{8\sqrt{3}\pi\omega}
\frac{1+v^2}{(1-v^2)^{1/3}}\,,
  \ee
where $v=(E_--E_+)/\omega$, which exhibits similar scaling and
universality features. The corresponding total probability reads
 \be
W=\frac{15 e^2 2^{2/3} \Gamma^2(2/3)\Gamma(5/6) (3eHE)^{2/3}}{28
\sqrt{3}\pi\omega}\,.
  \ee

\section{Conclusion}
To summarize: massless magnetized QED for spins $s=0,\, 1/2$
consistently describes synchrotron radiation and one-photon pair
creation of massless charges. In the quasiclassical regime (high
Landau levels of initial and final states) these  processes exhibit
common scaling law $(eHE)^{2/3}$ and have  universal spectral
distributions not depending on any parameters.

It is amusing to note that just during the  MG-14 conference the
announcement came about discovery of massless charged {\em
quasi-particles}  in semi-metals \cite{Xu:2015cga}.

\section*{Acknowledgements}
The author is grateful to  V.~G.~Bagrov, V.~A.~Bordovitsyn   and
A.~V.~Borisov for useful communications. This work was supported by
the Russian Foundation for Fundamental Research under Project
14-02-01092.


\begin{thebibliography}{0}

\bibitem{LL}
  L.~D.~Landau and E.~M.~Lifshitz,
  Addison-Wealey, Reading, Mass., 1971.

\bibitem{Kosyakov:2007pm}
  B.~P.~Kosyakov,
  J.\ Phys.\ A {\bf 41}, 465401 (2008)
  [arXiv:0705.1228 [hep-th]].

\bibitem{Lechner:2014kua}
  K.~Lechner,
  J.\ Math.\ Phys.\  {\bf 56}, no. 2, 022901 (2015)
  [arXiv:1405.4805 [hep-th]].

\bibitem{Vaks:1961}
  V.~G.~Vaks,
  Soviet\ Physics\ JETP  {\bf 13} 556-561 (1961).

\bibitem{Case:1962zz}
  K.~M.~Case and S.~G.~Gasiorowicz,
  Phys.\ Rev.\  {\bf 125}, 1055 (1962).

\bibitem{Kinoshita:1962ur}
  T.~Kinoshita,
  J.\ Math.\ Phys.\  {\bf 3}, 650 (1962).

\bibitem{Lee:1964is}
  T.~D.~Lee and M.~Nauenberg,
  Phys.\ Rev.\  {\bf 133}, B1549 (1964).

\bibitem{Weinberg:1965nx}
  S.~Weinberg,
  Phys.\ Rev.\  {\bf 140}, B516 (1965).

\bibitem{Fomin76}
P.~I.~Fomin, V.~A.~Miranski and Y.~A.~Sitenko,
Phys. Lett. {\bf 64B}, 444 (1976).

\bibitem{Gribov:1981jw}
  V.~N.~Gribov,
  Nucl.\ Phys.\ B {\bf 206}, 103 (1982).

\bibitem{Morchio:1985re}
  G.~Morchio and F.~Strocchi,
  Annals Phys.\  {\bf 172}, 267 (1986).

\bibitem{Ferrer:2012pb}
  E.~J.~Ferrer, V.~de la Incera and A.~Sanchez,
  Nucl.\ Phys.\ B {\bf 864}, 469 (2012).


\bibitem{Gal'tsov:2015cla}
 D.~V.~Gal'tsov,
  Phys.\ Lett.\ B {\bf 747}, 400 (2015).


\bibitem{Sokolov1}
   A.~A.~Sokolov and I.~M.~Ternov, Synchrotron Radiation, Pergamon, New
   York, 1968.

\bibitem{Bor}
``Synchrotron Radiation Theory and Its Development,''
V.~A.~Bordovitsyn (ed.),    World Scientific, Singapore, New Jersey,
London, 1999, 447 pp.

\bibitem{Schwinger:1973}
   J.~Schwinger, ``Particles, Sources, and Fields'',
   Addison-Wesley, Heading, Mass., Vol. III, Chap. 5, Sec. 6;
   Phys. Rev. D 7, 1969 (1973).

\bibitem{Tsai:1974cc}
  W.~Y.~Tsai,
  Phys.\ Rev.\ D {\bf 8}, 3460 (1973).

\bibitem{Klepikov}
N.~P.~Klepikov, Zh.Eksp.Teor.Fiz., {\bf 26}, 19 (1954).

\bibitem{TE}
W. Tsai and T.Erber,
Phys.Rev., D10, 492 (1974);\\
  V.~N.~Baier and V.~M.~Katkov,
  Phys.\ Rev.\ D {\bf 75}, 073009 (2007).

\bibitem{Xu:2015cga}
  S.~Y.~Xu {\it et al.},
  Science vol. 349 no. 6248 pp. 613-617 (2015).

\end{thebibliography}
\end{document}